\documentclass[aps, reprint, 10pt, superscriptaddress, prb]{revtex4-1}
\usepackage[version=3]{mhchem}
\usepackage{float}
\usepackage{longtable}
\usepackage{mathtools}
\usepackage[utf8]{inputenc}
\usepackage[T1]{fontenc}
\usepackage{lmodern}
\usepackage{url}
\usepackage{color}
\usepackage[table]{xcolor}
\usepackage{cleveref}

\begin{document}
\title{Efficient Charge Separation in 2D Janus van der Waals Structures with Build-in Electric Fields and Intrinsic p-n Doping}

\author{Anders C. Riis-Jensen}
\affiliation{CAMD, Department of Physics,
Technical University of Denmark, DK - 2800
Kongens Lyngby, Denmark}
\affiliation{These authors contributed equally to this work}
\author{Mohnish Pandey}
\affiliation{CAMD, Department of Physics,
Technical University of Denmark, DK - 2800
Kongens Lyngby, Denmark}
\affiliation{These authors contributed equally to this work}
\author{Kristian S. Thygesen}
\affiliation{CAMD, Department of Physics,
Technical University of Denmark, DK - 2800
Kongens Lyngby, Denmark}
\affiliation{Center for Nanostructured Graphene (CNG),
Department of Physics, Technical University of Denmark, DK - 2800 Kongens Lyngby, Denmark}
\email{thygesen@fysik.dtu.dk}
\date{\today}

\begin{abstract}
Janus MoSSe monolayers were recently synthesised by replacing S by Se on one side of MoS$_2$ (or vice versa for MoSe$_2$). Due to the different
electronegativity of S and Se these structures carry a finite out-of-plane dipole moment. As we show here by means of density functional theory
(DFT) calculations, this intrinsic dipole  leads to the formation of built-in electric fields when the monolayers are stacked to form $N$-layer structures. For sufficiently thin structures ($N<4$) the dipoles add up and shift the vacuum level on the two sides of the film by $\sim N \cdot 0.7$ eV. However, for
thicker films charge transfer occurs between the outermost layers forming atomically thin n- and p-doped electron gasses at the two surfaces. The doping concentration can be tuned between about $5\cdot 10^{12}$ e/cm$^{2}$ and $2\cdot 10^{13}$ e/cm$^{2}$ by varying the film thickness. The surface charges counteract the static dipoles leading to saturation of the vacuum level shift at around 2.2 eV for $N>4$. Based on band structure calculations and the Mott-Wannier 
exciton model, we compute the energies of intra- and interlayer excitons as a function of film thickness suggesting that the Janus multilayer films are ideally suited for achieving ultrafast charge separation over atomic length scales without chemical doping or applied electric fields. Finally, we explore a number of other potentially synthesisable 2D Janus structures with different band gaps and internal dipole moments. Our results open new opportunities for ultrathin opto-electronic components such as tunnel diodes, photo-detectors, or solar cells.       
\end{abstract}

\maketitle
\section{Introduction}
The unique optical properties of atomically thin crystals combined with the possibility to combine them into lateral and vertical heterostructures have 
placed two-dimensional (2D) materials at the forefront of photonic and optoelectronic materials research. Among the unique optical properties that distinguish the 2D materials from the more conventional bulk semiconductors are their strong light-matter
interactions\cite{britnell2013strong,splendiani2010emerging,mak2010atomically} and pronounced excitonic effects\cite{ye2014probing,klots2014probing}. Furthermore By stacking individual 2D materials into van der Waals (vdW) heterostructures~\cite{geim2013van} their optical 
properties can be further controlled by engineering of the band structure~\cite{withers2015light} or the dielectric
environment\cite{andersen2015dielectric,gjerding2017layered}. 

While the strong excitonic effects in 2D semiconductors are of interest for some applications they can pose a serious problem for others. This holds 
in particular for photodetectors and solar cells, which rely on efficient dissociation of photo-excited excitons into free electrons and holes. For 
in-plane charge separation, the problem has been overcome by forming lateral pn-junctions using split gate
techniques\cite{massicotte2018dissociation,ross2014electrically} which creates a sufficiently large potential gradient to dissociate the
excitons\cite{haastrup2016stark}. For out-of-plane device architectures, the exciton dissociation has been achieved using hetero-bilayers, e.g. MoS$_2
$-WSe$_2$, with natural Type-II band alignment\cite{peng2016ultrafast,chen2016ultrafast} or by applying an external bias voltage across a $N$-layer 
stack, e.g. five layers of WSe$_2$~\cite{massicotte2016picosecond}. 

Here we propose a novel type of vdW bonded $N$-layer structure with an intrinsic electric field in the out-of-plane direction stemming from 
an out-of-plane asymmetry and finite dipole moment of the individual monolayers. Above a certain critical thickness the built-in electric field becomes compensated by surface charges accumulating at the surfaces leading to natural n- and p-doping of the two outermost monolayers thus generating an ultrathin pn-junction. The electric field strength, the electronic band alignment throughout the structure, and the doping concentration at the surface layers, can be tuned 
to some extent by varying the film thickness. We show that the build-in electric field in structures with up to around 20 layers is sufficient to dissociate intralayer excitons into interlayer excitons, which is the critical step for achieving charge separation. Finally, we show that these unique properties are not limited to MoSSe. In fact, our first-principles calculations predict a number of other stable and potentially synthesisable 2D materials with finite dipole moments. By stacking different types of 2D Janus structures one could potentially engineer, not only the band edge positions but also the internal electric field and the doping concentration at the surface layers.     

\section{Methods}
All calculations have been performed with the GPAW~\cite{enkovaara2010electronic} code. The in-plane lattice constant of the monolayer
MoSSe is calculated with the PBE functional.\cite{PhysRevLett.77.3865}. In order to get an accurate description of the interlayer
distance, we use the BEEF-vdW functional which includes the non-local van der Waals correction.\cite{PhysRevB.85.235149}
We find that the calculated interlayer distance
does not change as the number of layers in the Janus structure is increased from 2 to 3. Therefore, we take the interlayer distance calculated for the
bilayer structure as the optimum distance for all the multi-layer structures. The wave functions are expanded in a plane wave
basis with an energy cutoff of 800 eV. For structural relaxations we employ a 18$\times$18$\times$1 Monkhorst-Pack grid.\cite{PhysRevB.13.5188}
The PBE band structures are calculated on very fine a 54$\times$54$\times$1 $k$-point grid with a 800 eV 
plane wave cutoff and inclusion of spin-orbit coupling. A vacuum region of 15 \AA\ is inserted in the perpendicular direction to separate the periodically repeated images. A Fermi smearing of 0.01 eV was used for all the calculations. 

In order to calculate the screened electron-hole interaction in the multilayer structures, the dielectric
building block of monolayer MoSSe is calculated following Ref. \citenum{andersen2015dielectric} to be used as input for the Quantum Electrostatic Heterostructure (QEH) model. The calculations are performed in the random phase approximation using wave functions and eigenvalues from a PBE ground state calculation with a 100$\times$100$\times$1 $k$-point grid and 
800 eV plane wave cutoff. For the density response function a plane wave cutoff of 150 eV is used (to account for local field effects).  The in-plane exciton effective mass used in the 2D Mott-Wannier model is calculated from 
the PBE band structure and is found to be $\mu_\textrm{ex} = 0.24 m_0$ (for the direct exciton at the $K$-point), where $m_0$ is the free space electron mass. The dielectric constant for bulk MoSSe (in which MoSSe layers are stacked together as in 2H MoS$_2$ bulk) was obtained using random phase approximation (RPA) based on a PBE ground state calculation. The ground state calculation of the bulk MoSSe to be used as a starting point for the RPA calculation
was done on a $24\times 24 \times 18$ $k$-point grid using a 800 eV plane wave cut-off. The number of bands was set to six times the number of valence bands and we converged 5 times the number of valence bands in the groundstate calculation. All of these bands were used in the RPA calculation, which had a plane wave cut-off of 125 eV. \\

For more details on the calculation of intra- and interlayer exciton binding energies from the 2D Mott-Wannier model we refer to the detailed accounts given in Ref. \onlinecite{Latini2015} and \onlinecite{latini2017interlayer}.

\section{Results and Discussions}
Janus MoSSe monolayers were recently synthesized using both controlled sulfurization of MoSe$_2$\cite{zhang2017janus} and selenization of
MoS$_2$\cite{Lu2017janus}. Following the experimental realizations, a number of computational studies have considered various aspects of MoSSe 
monolayers including magnetism\cite{meng2018ferromagnetism,wang2018tuning} as well as electronic, optical, and transport
properties\cite{yin2018tunable,li2017electronic,ma2018janus}. One study also explored multilayers of MoSSe and reported an observed rapid closing of 
the band gap as function of the number of layers saturating at a value around 0.1 eV for $N>3$, but without providing a physical explanation for this 
effect\cite{guan2018tunable}. In contrast to these findings, we show that the Janus multilayers undergo an insulator to metal transition at around $N=4$, 
which is driven by the internal dipole of the structure.  

\begin{figure*}
\includegraphics[width=\textwidth, height=!]{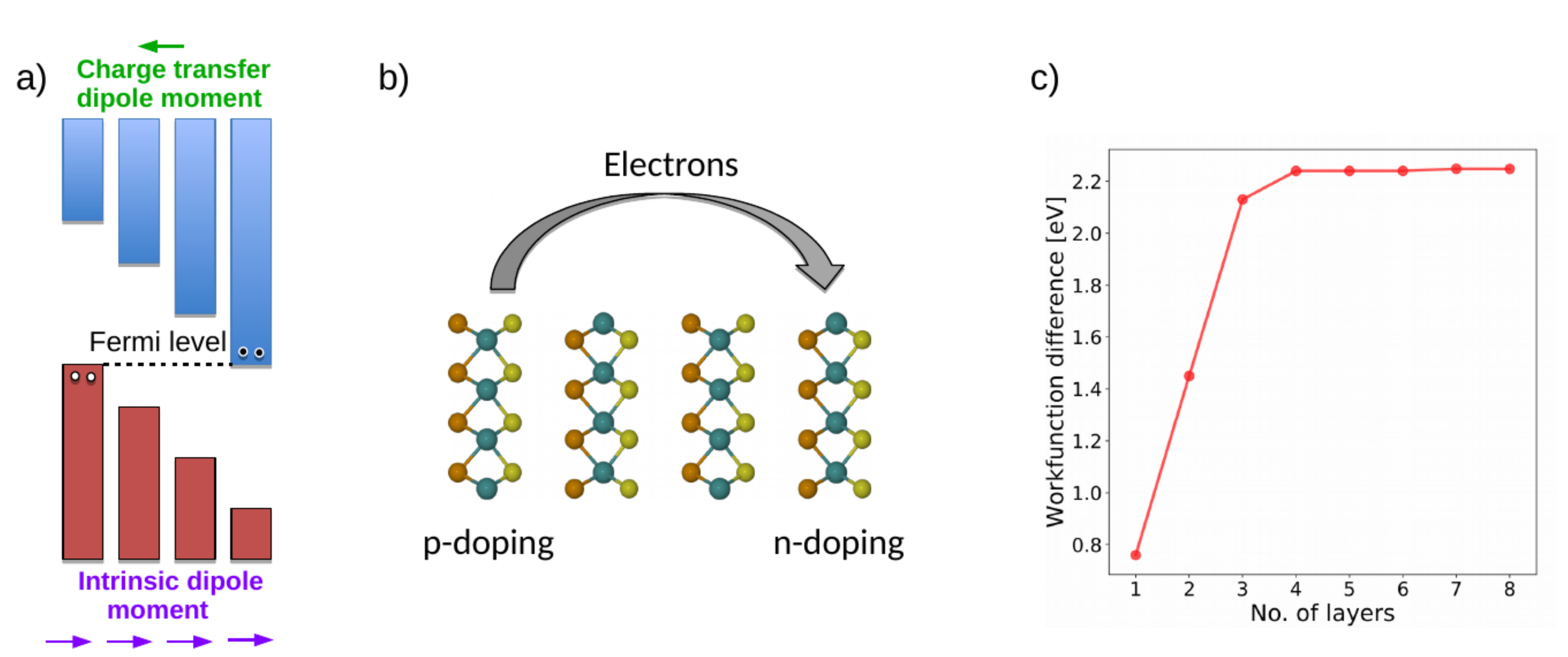}
\caption{a) Qualitative sketch of the charge transfer at four layers with the intrinsic dipole moment shown below and the dipole moment due to the charge separation above. b) sketch of the p- and n-doping of the outermost layers. c) Difference between the workfunction on either side of a $N$-layer Janus structure calculated from the PBE bandstructure. }\label{MoSSe_wf_diff}
\end{figure*}

To obtain the equilibrium structure we have performed density functional theory (DFT) calculations for AB stacked MoSSe multilayer structures using 
the BEEF-vdW exchange-correlation functional\cite{PhysRevB.85.235149} as implemented in the GPAW electronic structure
code\cite{enkovaara2010electronic} (see Methods for more details). We find in-plane lattice constant of 3.251 \AA\ in good agreement with previous 
work\cite{ji2018janus,zhang2017janus,Lu2017janus} and interlayer spacings of 6.896 \AA\ (defined as the Mo-Mo distance), which is found to be practically independent of the number of layers in the multilayer structure. An example of a 4 layer 
structure is shown in figure \ref{MoSSe_wf_diff}b. 

\begin{figure*}
\includegraphics[width=\textwidth, height=!]{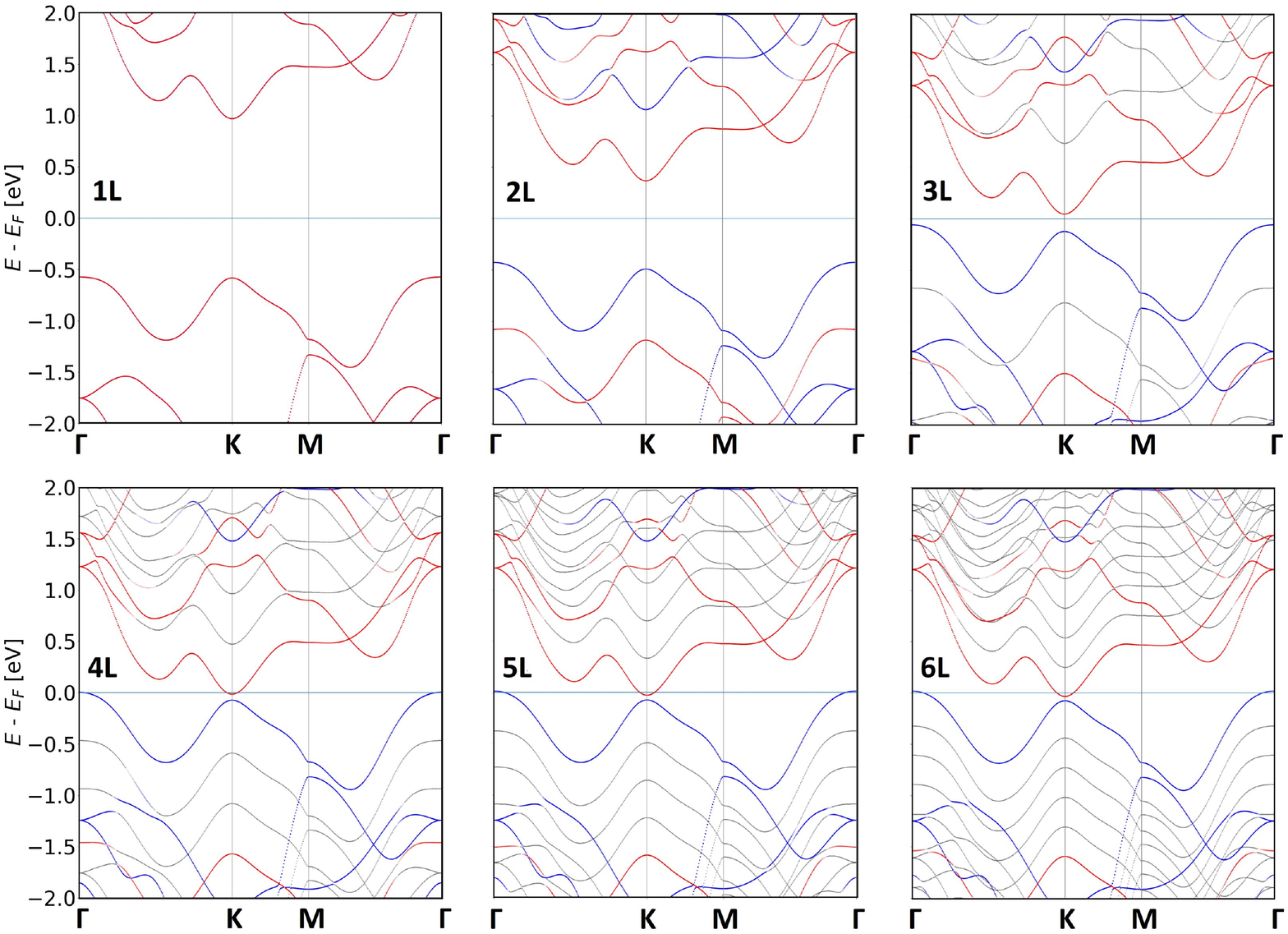}
\caption{Bandstructures for the 1-6 layer Janus structure. Red and blue shows the bands projected onto the top and bottom layers respectively, while 
all layers in between are colored gray. Faded blue or red indicates a high hybridzation between neighbouring layers. From there one can see a direct 
band gap for the monolayer MoSSe at the K point, while all multilayer structures have an indirect band gap from $\Gamma$ to K, with the valence band 
maximum and conduction band minimum located at the bottom and top layer respectively.}\label{MoSSe_All_BS_color}
\end{figure*}

The difference in electronegativity of sulfur and selenium leads to the formation of a static
dipole of 0.038 |e|\AA\ per lateral unit cell across each MoSSe monolayer. When several layers are stacked together, these dipoles add up and generate a potential gradient in the direction perpendicular to 
the film. The evolution of the band structures of MoSSe multilayer structures for $N=1-6$ are shown in figure \ref{MoSSe_All_BS_color}. The band structures are 
obtained with the PBE functional\cite{perdew1996generalized}. Interestingly, the band structures change dramatically with $N$ despite of the fact 
that wave functions on neighbouring layers hybridize only weakly due to the weak vdW bonds. For $N \geq 4$ the band gap vanishes and the Fermi level 
intersects the valence band maximum (VBM) at the $\Gamma$-point and the conduction band minimum (CBM) at the $K$-point of the lateral 2D Brillouin zone. The bands 
located on the top and bottom layer are colored red and blue, respectively. Clearly, the charge transfer occurs between the outermost layers of the structure.

The insulator-metal (IM) transition occurs when the built-in potential difference between 
the top and bottom layers exceeds the band gap, $E_G$. At this point the CBM moves below the VBM resulting in a net positive (negative) charge in the top (bottom) layer. This shift in the charge density creates a dipole in the direction opposite to the intrinsic dipoles of the MoSSe layers. The effect is clearly visible from the difference in the electrostatic potential on the two sides of the 
multilayer structure, see figure \ref{MoSSe_wf_diff}c. In the same figure we show a qualitative sketch of the charge transfer, the intrinsic dipoles of the individual monolayers, and the counter balancing dipole due to the charge transfer (a). The potential difference across an $N$-layer structure, $\Delta \Phi(N)$, is seen to saturate at a value around 2.2 eV. Adding more layers 
will increase the potential difference created by the internal dipoles (which simply add up). However, after the onset of the IM transition any increase in the internal dipole will be counterbalanced by surface charges transferred between the outermost monolayers. The number of layers at which this insulator-metal transition occurs is approximately  
\begin{equation}
N_{\mathrm{IM}}=E_G/\Delta \Phi_0+1,
\end{equation}
where $\Delta \Phi_0$ is the potential difference across a single layer. For the MoSSe monolayer we find $\Delta \Phi_0=0.76$ eV and $E_G = 1.54$ eV, yielding $N_{\mathrm{IM}}=3$, in good agreement with the band structure calculations in figure \ref{MoSSe_All_BS_color}. We note in passing that, due to the well known underestimation of the band gap by the PBE xc-functional, $N_{\mathrm{IM}}$ might also be underestimated. In fact, using the G$_0$W$_0$ calculated band gap for monolayer MoSSe of 2.33 eV, we obtain $N_{\mathrm{IM}}=4$. 

For $N>N_{\textrm{IM}}$ the charge density at either of the two surface layers can be estimated from a simple plate capacitor model, 
\begin{equation}\label{eq:charge}
\sigma(N) = \frac{\epsilon_{\bot} \Delta \Phi_0}{d} \left(1 - \frac{N_{\mathrm{IM}}}{N} \right) ,
\end{equation}    
where $\epsilon_{\bot}$ is the dielectric constant of bulk MoSSe in the out-of-plane direction, $\Delta \Phi_0$ is the potential difference created by a single layer, and $d$ is the interlayer distance. We have calculated the dielectric constant within the random phase approximation (RPA) for bulk MoSSe and obtain $\epsilon_{\bot} = 3.68 \epsilon_0$ (see methods section for further details). The prefactor in eq. (\ref{eq:charge}) corresponding to the charge density in the limit $N \rightarrow \infty$, then becomes  $\sigma(\infty) = 2.3 \cdot 10^{13}$ e/cm$^2$. 

We next turn to an analysis of the charge separation ability of the Janus structures.
Upon light illumination, electron-hole (e-h) pairs will be generated within the structure. Due to the small spatial overlap of the wave functions in neighbouring 
layers, the generated e-h pairs will predominantly be of the intralayer type. The photoexcited e-h pairs will thermalise rapidly on a sub-pico second timescale\cite{nie2014ultrafast}. In comparison e-h recombination in similar TMD structures \emph{without} build-in electric fields occurs on time scales of at least several ps\cite{palummo2015exciton,shi2013exciton,lagarde2014carrier}. At room temperature, the e-h recombination is dominated by defect assisted processes and consequently significantly longer e-h lifetimes are expected for highly pure samples. After thermal relaxation, the resulting non-equilibrium distribution includes hot electrons and holes with energies above the band gap as well as bound excitons. The hot electrons and holes will separate efficiently in the large build-in electric field. Consequently, we focus on the bound excitons which are more difficult to dissociate. 

Excitons in layered TMD structures have binding energies in the range 0.5 eV (isolated monolayers) to 0.1 eV (bulk), which are significantly larger than $k_BT$ at room temperature. The crucial first step of the exciton dissociation process is the transformation 
of the intralayer excitons into an interlayer exciton with the electron and hole located on neighbouring layers. This process requires that the energy of the interlayer exciton is equal to or lower than the intralayer exciton. In the absence of an electric field, this condition is never satisfied because of the weaker e-h binding energy in the spatially separated interlayer exciton\cite{deilmann2018interlayer}. However, in structures with a built-in electric field this energy difference can be overcome by the band offset between neighbouring layers. 

To determine the conditions for exciton dissociation, we calculate the binding energies of intra- and interlayer excitons in stacked MoSSe as a function of film thickness. We use a 2D Mott-Wannier model, which has been shown to yield accurate binding energies for excitons in mono- and few-layer TMDs\cite{Cudazzo2011,Pulci2012,Latini2015} as well as 
for interlayer excitons in vdW heterostructures\cite{latini2017interlayer}. The 2D Mott-Wannier Hamiltonian takes the form

\begin{equation}
\label{eq:MW}
\left[-\frac{\nabla_{2D}^2}{2\mu_\textrm{ex}}+W({\bf r}_\parallel)\right]F({\bf r}_\parallel)=E_\textrm{b}F({\bf r}_\parallel),
\end{equation}

where $\mu_\textrm{ex}$ is the exciton effective mass and $W({\bf r}_\parallel)$ is the electron-hole interaction energy. The exciton effective mass 
is defined as $\mu_\textrm{ex}^{-1}=m_\textrm{e}^{-1}+m_\textrm{h}^{-1}$, where the hole and electron masses must be evaluated at the band extremum of the relevant layer. Assuming direct (i.e. zero-momentum) excitons, both the electron and hole masses should be evaluated at the $K$-point of the BZ yielding an exciton mass of $\mu_\textrm{ex} = 0.24 m_0$. We stress that Eq. (\ref{eq:MW}) remains valid in the case of interlayer 
excitons because, even though the electron and the hole are now spatially separated in the out-of-plane direction, 
their motion is still confined to their respective layers. On the other hand, this spatial separation affects the screened electron-hole interaction 
$W$. We calculate $W$ using the Quantum Electrostatic Heterostructure (QEH) model\cite{andersen2015dielectric} to include the additional screening from the surrounding layers. We obtain binding energies for intralayer excitons in the center of the film in the range of 0.38 eV (for $N=3$) to 0.23 (for $N\to \infty$) and interlayer exciton binding energies from 0.29 eV (for $N=3$) to 0.15 (for $N\to \infty$).

\begin{figure*}[!]
\includegraphics[width=12cm, height=!]{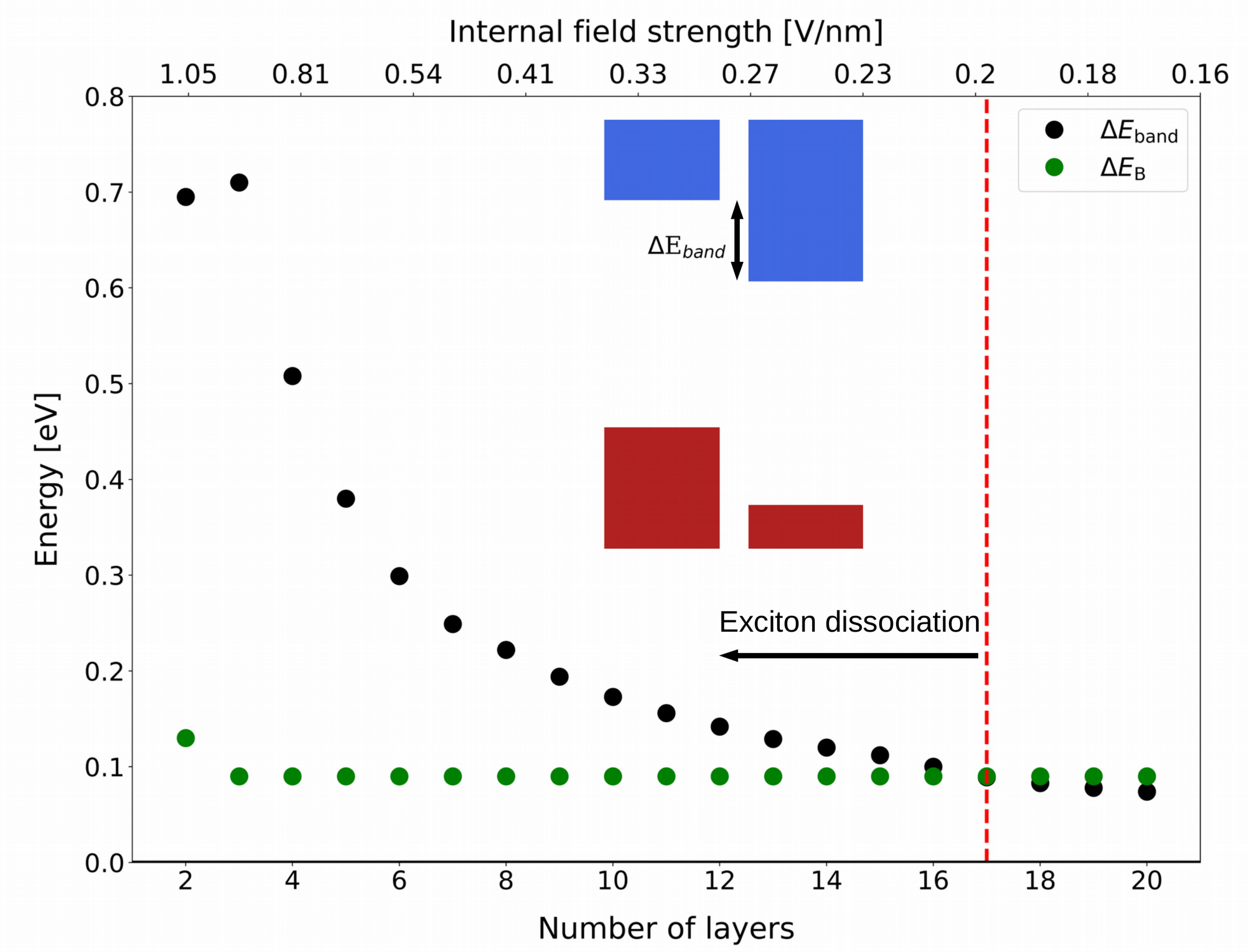}
\caption{Difference in intralayer exciton binding energy ($E_B^{intra}$) and interlayer exciton binding energy ($E_B^{inter}$) in green for the central layer and band offset between neighbouring layers in black also for the central layer in a $N$ layer structure.  At the crossover between the two curves at around $N = 17$ layers, the interlayer exciton is no longer energetically favourable and exciton dissociation cannot take place. This critical limit is shown by the red dashed line. The inset shows the definition of the band offset between neighbouring layers.}\label{Janus_excitons2}
\end{figure*}

In figure \ref{Janus_excitons2} we show the difference in binding energy between the intralayer and interlayer excitons in the central layers of an $N$-layer MoSSe structure (green symbols)
\begin{equation}
\Delta E_B(N)= E_B^{\mathrm{intra}}(N)-E_B^{\mathrm{inter}}(N)
\end{equation}
The black curve shows the band offset between two neighbouring layers of the structure, as shown in the inset. The two curves cross around $N=17$ indicated by the red dashed line. For structures thicker than this critical thickness, the difference in exciton binding energy cannot be overcome by the band offset and the exciton cannot dissociate (more precisely, the driving force for exciton dissociation is strongly reduced). 

The above analysis is based on a picture where excitons are composed of electrons and holes bound to specific layers. Alternatively, we can describe the excitons by a homogeneous anisotropic 3D Mott-Wannier model where the layered nature of the Janus structure is accounted for by using different dielectric constants and effective masses in the in- and out-of-plane directions. Such a model was developed in Ref. \onlinecite{pedersen2016exciton} and applied to bulk TMDs. Using input parameters from first-principles calculations the model yields binding energies, $E_B$, of 83 meV and 52 meV for bulk MoS$_2$ and MoSe$_2$, respectively, and exciton radius in the out-plane directions ($a_0^*$) of around 1.1 nm and 1.4 nm. Assuming similar values for bulk MoSSe, the characteristic field strength at which this exciton dissociates, $E_B/a_0^*$, becomes roughly 0.1 V/nm. It can be seen that the result for the critical thickness obtained with this homogeneous anisotropic 3D model is in reasonable agreement with the result obtained with the layered 2D exciton model. In fact, by extrapolating the results we find the internal field strength to reach 0.1 V/nm at around 30 layers. We stress that according to Ref. \onlinecite{pedersen2016exciton} the exciton dissociation rate for out-of-plane field strengths of 0.1 V/nm is well above $10^{13}$ s$^{-1}$ for both direct and indirect excitons in both MoS$_2$ and MoSe$_2$. From this we conclude that exciton dissociation in MoSSe Janus structures up to the critical thickness of 17 layers occurs much faster than the exciton recombination which is characterised by rates $<10^{12}$ s$^{-1}$.\cite{palummo2015exciton,shi2013exciton,lagarde2014carrier}. 

\begin{figure}
\includegraphics[width=8cm, height=!]{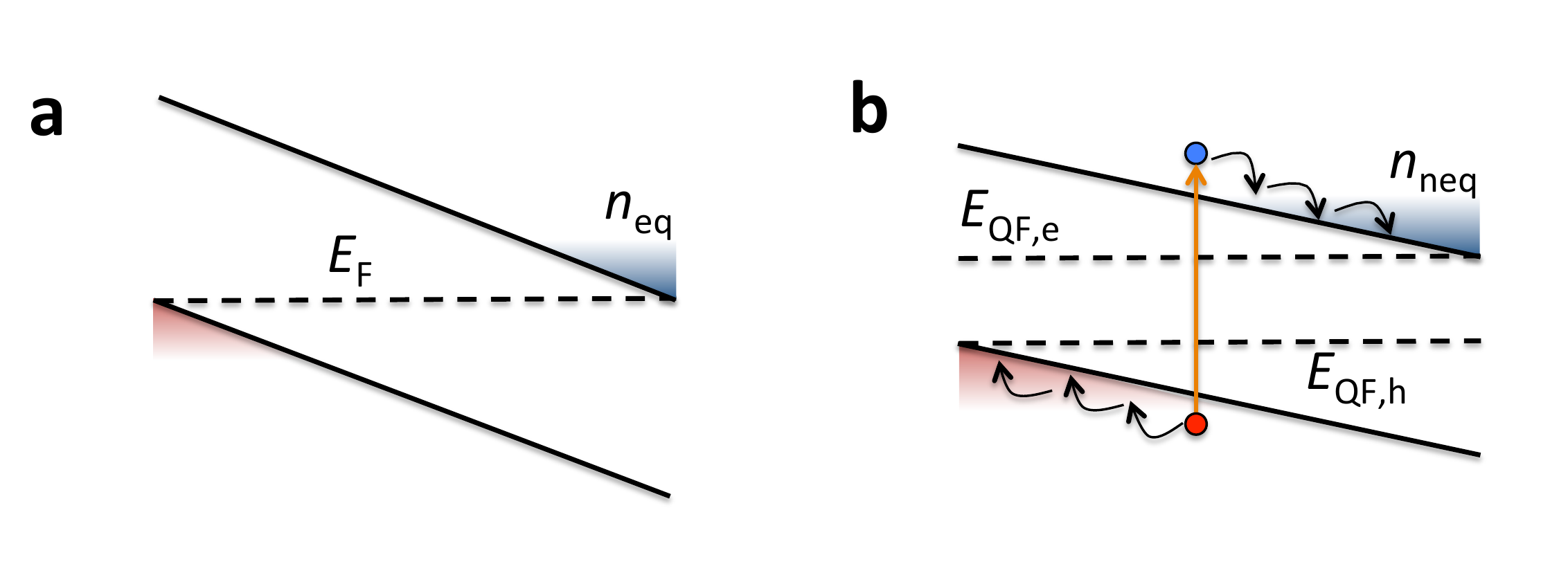}
\caption{Sketch of the dissociation of excitons. In a) is shown the band alignment at equilibrium, while b) shows the exciton dissociation after photo-excitement and the corresponding quasi-Fermi levels. The spatial separation of the holes and the photo-excited electrons sets up an opposite dipole, which again opens a band gap.}\label{janus_noneq}
\end{figure}

In this paper we have focused on the charge separation ability of MoSSe Janus structures. This property is essential for a number of opto-electronic devices including photodetectors and solar cells. The latter application might seem impossible considering the band structures in figure \ref{MoSSe_All_BS_color} which shows a decreasing band gap reaching zero for $N>3$. However, one should keep in mind that this is the situation in equilibrium. Upon excitation charge carriers excited in the interior of the pn-junction will separate due to the build-in field and electrons (holes) will move to the n (p) side of the structure. This charge imbalance will create a dipole opposite to the build-in field (just like the charge transfer creating the p and n surface doping in equilibrium). The size of the non-equilibrium charge distributions will determine the achievable photo-voltage which is given by the difference in the (quasi-)Fermi levels of electrons and holes, respectively. The size of the non-equilibrium charge distributions, and thus the achievable voltage, will depend on the carrier lifetimes (limited by recombination processes) relative to the rate of charge separation. Assuming a conservative recombination rate of 1 ps, the critical field strength at which perpendicular exciton dissociation in bulk TMDs dominates the recombination is 0.01 V/nm\cite{pedersen2016exciton}, which is easily achieved in the stacked Janus structures, cf figure \ref{Janus_excitons2}. We conclude that it should be possible to realize finite photo-voltages in stacked Janus structures, even for films with $N>N_{\mathrm{IM}}$. The situation is illustrated in figure \ref{janus_noneq}.

We mention that other applications of (stacked) 2D Janus structures could be envisioned such as the tunnel diodes, e.g. to separate light absorbing layers in multi-junction solar cells, tunnel field effect transistors\cite{sarkar2015subthermionic}, or for tuning band alignment or Schottky barriers in van der Waals heterostructures\cite{lee2018two}.

\begin{figure}
\includegraphics[width=8.2cm, height=!]{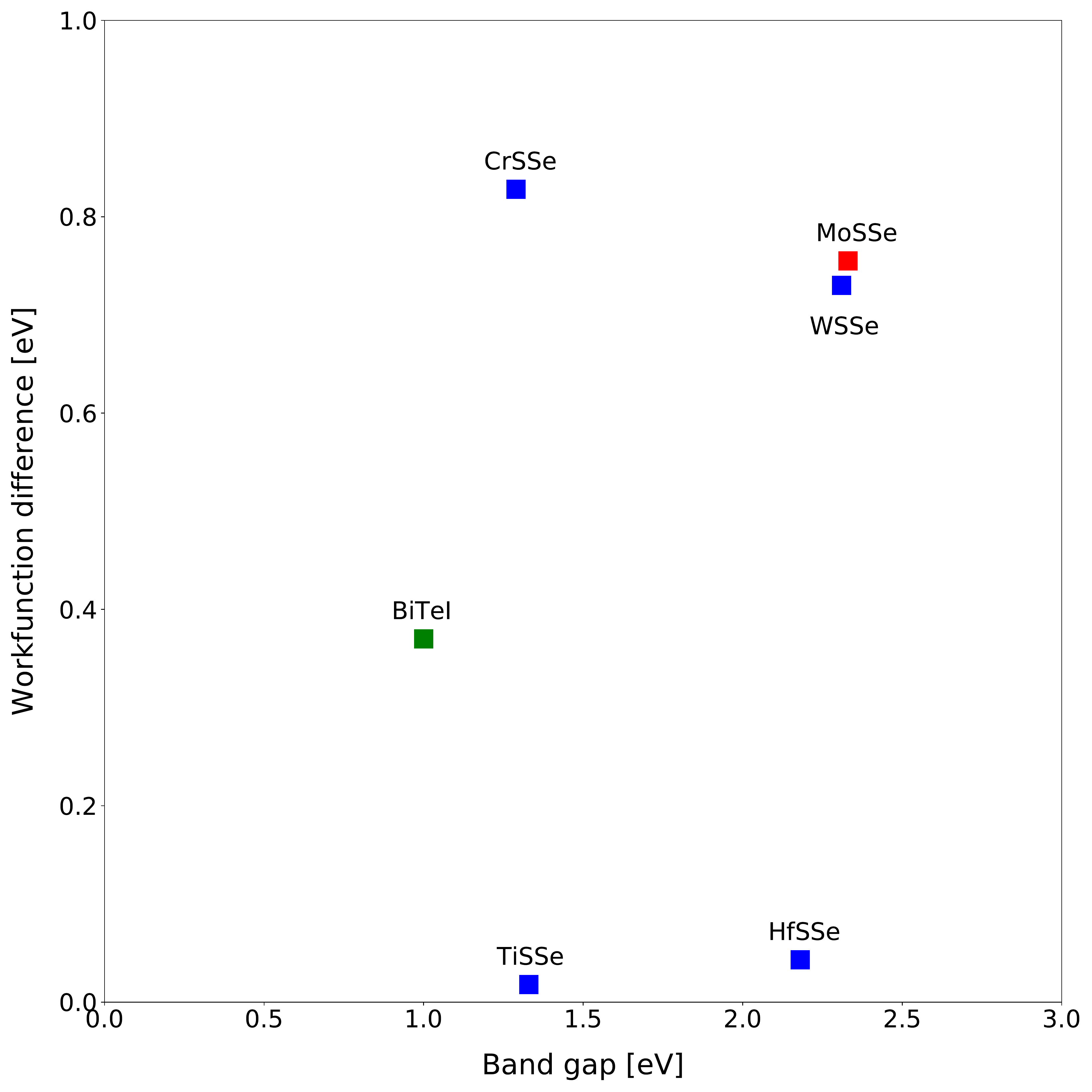}
\caption{Workfunction difference and G$_0$W$_0$ band gap for five different TMD monolayer Janus structures with MoSSe, the structure considered in this study highlighted in red, and in green BiTeI which has a different structure than MoSSe and the other 4 structures in blue which all share the same structure as MoSSe. This shows that the charge transfer and band alignment shift can also be controlled by the material in addition to adjusting the number of layers. }\label{workdiff}
\end{figure}

Finally, we show that finite out-of-plane dipole moments in 2D materials are not limited to the MoSSe monolayer. We have performed DFT calculations for a number of monolayers with similar structures and chemical compositions as MoSSe. The results of these calculations, including the atomic structure, total energies, electronic band structure and much more, are directly available in the Computational 2D Materials Database (C2DB)\cite{c2db}. For the monolayers found to be both dynamically and thermodynamically stable (according to the criteria used in the C2DB and described in Ref. \citenum{c2db} we show in figure \ref{workdiff} the relation between the G$_0$W$_0$ band gap and the work function difference. We find a linear relation between the work function difference and the internal dipole moment for the 5 structures sharing the same geometry as MoSSe. We here therefore only show the work function difference, which is the interesting parameter for an actual experimental realization. We note in passing that BiTeI is known as layered bulk material and should be easily exfoliable according to Ref. \citenum{mounet2018two}. Indeed, BiTeI was recently exfoliated and studied in its monolayer form\cite{ma2014emergence}. Returning to figure \ref{workdiff} we observe a large variation in the key electronic properties of these Janus structures. This suggests that in addition to controlling the number of layers in the Janus structure, it is also possible to control the size of the build-in field and therefore the surface layer doping level, by varying the type of material. In particular, by combining different 2D Janus layers into van der Waals heterostructures, it should be possible to design not only the band alignment but also the build-in electric field and e.g. obtain non-linear potential profiles. 

\section{Conclusion}
In summary, our first-principles calculations show that 2D vdW structures consisting of stacked Janus MoSSe monolayers host a strong built-in 
electric field of about 0.1 V/\AA. The electric field induces a staggered band alignment throughout the structure, and at a critical thickness of 3-4 
layers, the CBM of the top layer meets the VBM of the bottom layer triggering an electron transfer between the outermost monolayers. The charge 
density on the surface layers of this natural vertical p-n junction can be tuned between about
$5 \cdot 10^{12}$ e/cm$^2$ and $2 \cdot 10^{13}$ e/cm$^2$ 
by varying the film thickness (these values corresponds to the cases of 5 and 17 layers, respectively). Using many-body GW calculations in combination with a 2D Mott-Wannier model we estimated the energy of intralayer and 
interlayer excitons as function of film thickness. These calculations show that for film thickness below approximately 17 layers, the shift in band edges at 
neighbouring layers exceeds the difference in binding energy of the intra- and interlayer excitons and thus facilitates the spontaneous dissociation 
of photo-generated intralayer excitons into spatially separated electron-hole pairs. Based on these results we propose that Janus vdW structures 
could be used as basis for ultrafast and ultrathin photodetectors or electrical diodes. In the future, it would be interesting to explore the 
possibility of using 2D Janus structures to introduce highly local potential gradients in intrinsic semiconductors without the need for doping. This 
could be interesting for charge separation in solar cells or for tuning band alignment at the interface between two different semiconductors.  

\section*{acknowledgement}
This project has received funding from the European Research Council (ERC)
under the European Union's
Horizon 2020 research and innovation programme (grant agreement No 773122,
LIMA). The Center for Nanostructured Graphene is sponsored by the
Danish National Research Foundation, Project DNRF103. \\   

\newpage
\bibliography{janus_refs}

\end{document}